The effect of separatrix density and PFC material on H-mode confinement in the ITPA global H-mode database

M. Kotschenreuther[1], X. Liu[1], D.R. Hatch[2,1], S. M. Mahajan[2,1]

[1]ExoFusion
[2]University of Texas at Austin

Abstract

Recent data, added to the ITPA global H-mode database [1] for ASDEX-U [2] and JET-ILW, reveals that the separatrix density $n_{sep}$ has a correlation with the H20 factor (Confinement time relative to the ITPA20-IL scaling) [1]. These trends are analyzed in detail. They are not a result of proximity to the density limit. The normalized $n_{sepN} = n_{sep}/\bar{n}$ is introduced, motivated by theory ($\bar{n}$ is the average density). The trends in $n_{sepN}$ can be understood in terms of the two main mechanisms of pedestal characteristics – MHD stability and recently developed theories of gyrokinetic transport. Careful analysis shows these mechanisms can be distinguished in the data. The most dramatic improvement in confinement time arises primarily from reductions in pedestal transport. A new definition of density peaking that includes core peaking is found to best explain H20 when advanced H-modes are included: $n_{sepN0} = n_{sep}/n(0)$, the inverse of the total density peaking from the separatrix to the axis. The highest H-factors are reached by the confluence of relatively low normalized $n_{sepN0}$ plus high Shafranov shift or poloidal beta. The importance of these two variables is also theoretically predicted from recent analysis of the gyrokinetic system, where a constraint can limit the access of ITG/TEM modes to free energy in equilibrium gradients. The Plasma Facing Component (PFC) material also shows a strong influence in the data. This is likely due to the importance of $n(0)/n_{sep}$ to attaining high H20, in conjunction with the known tendency for tungsten (W) to accumulate with density peaking and low transport. Preliminary results indicate that $n_{sepN0}$ might also be important with core ITBs.

Introduction

One of the most crucial and challenging issues in fusion is to integrate a high-performance core plasma with PFCs which must operate for long periods under very demanding conditions; it is imperative, therefore, to develop a better understanding of the effect of SOL conditions and PFCs on core confinement. Experiments on many devices do establish strong connection between global confinement and plasma conditions in the SOL and on the separatrix. For example, low Z wall conditioning is widely used to improve performance [3]. There is, in fact, considerable reported evidence that SOL or separatrix conditions are a determinant of plasma performance [4-17]: *Performance degradation* is conventionally attributed to high fueling, reduced pedestal MHD stability, approach to the density limit, or impurities, while *improvement* is attributed to reduced recycling or to improved pedestal MHD stability.

It is noteworthy that the observed improvements can be quite large, e.g., H98 (the confinement time $\tau_E$ relative to the ITER98(y,2) scaling) reached values up to ~ 1.8 on NSTX [7,8,13], up to

~ 2 for Li injection into the SOL on DIIID [9] [17] and up to ~2 with Li PFCs [6,16]. For some concepts of Fusion Power Plants, high H factors are very economically desirable [18,19]. And yet, parameters for SOL, separatrix and PFC material do not appear in most methods (including scaling laws) of predicting confinement. As a step to remedy this, data was added to the ITPA global H-mode database, where the electron separatrix density $n_{sep}$ was included [1] [2]. There is considerable variation of H-factor with $n_{sep}/n_G$ [1] ($n_G$ is the Greenwald density limit).

There are three fundamental fusion relevant areas that this paper attempts to explore:

1) First is to investigate, thoroughly, the confinements trends with $n_{sep}$; we will be fully guided by the contemporary theoretical understanding of H-mode confinement that clarifies the mechanisms by which $n_{sep}$ affects $\tau_E$. *Theory predicts that the pedestal density profile affects the pedestal stability and transport, which can be indicated by $n_{sep}/n_{ped}$. The data base results are consistent with this.*
2) Core density profiles are also expected to affect confinement. We consider these too.
3) Extract and categorize the $\tau_E$ dependence on the PFC materials.

Although the data is only 0D, it can still reveal a great deal: the existence of confinement trends as well as the likely origin of such trends. The data allows us to deduce, for instance, that some of the trends may result from changes in pedestal transport. Increased pedestal transport at high values of $n_{sep}/n_{ped}$ has been found to explain reduced confinement in JET-ILW [12][20-23]; here we find evidence of that trend in the opposite direction, namely, reduced pedestal transport at low of $n_{sep}/n_{ped}$ leading to improved confinement. This reinforces the importance of pedestal transport for determining H-mode confinement [24]. In addition, low $n_{sep}$ would be expected to correlate with low recycling; these results are consistent with the idea that improved core confinement with low recycling in the SOL influences the core by providing a low $n_{sep}$.

When advanced H-modes (called hybrids in this database) are included in the analysis, we will see that density peaking in the core can also be a significant variable to explain confinement. The single variable $n_{sep}/n(0)$ is found to best describe the trends in H20 (confinement time normalized to ITPA20-IL). The $n_{sep}/n(0)$ is the inverse of the *total* density peaking across the core, from the separatrix to the axis, and it includes density peaking arising in the pedestal and the core. This, in conjunction with a variable related to the poloidal beta $\beta_{pol}$ ($\alpha = \epsilon \beta_{pol}$), explains the trends in H20 rather well when both advanced H-modes and conventional ones are considered together in the data. Preliminary evidence finds that even ITB cases may be similar.

The emergence of these two crucial control variables, from the data base analysis, is remarkable because this is what is predicted by the by most recent gyrokinetic analysis of low transport regimes (that do not rely upon velocity shear) [25]. Due to a constraint in the gyrokinetic system, the turbulent transport cannot tap free energy in equilibrium gradients when the (logarithmic) density gradient is large and the local α is large. The global average of the theoretically relevant quantities across the discharge are α and $n_{sep}/n(0)$, the same two variables found to be most important in the data here. The theoretical considerations should apply equally to edge TBs (pedestals), ITBs and the core generally. For a very limited number of shots (where enough data is published) we have found that the variables α and $n_{sep}/n(0)$ appear to be important in cases with ITBs. We will discuss this further in the section on interpretation.

ITPA database analysis of the effects of separatrix density.

As an introduction, we plot the H factor (relative to the latest reference scaling ITPA20-IL [1]) versus the separatrix electron density normalized by the line averaged density $n_{sepN} = n_{sep}/\bar{n}$. ASDEX-U H-modes are shown in figure 1a, JET-ILW H-modes in Figure 1b, and ASDEX-U hybrid H-modes (referred to here as hybrids) in figure 1c. The wall materials and divertor configurations for ASDEX-U are of several types; wall materials including all C, all W, and mixtures of C and W. A regression line, with its associated 95% confidence intervals, is shown for each case. All show a trend toward higher H at lower $n_{sepN}$. At $n_{sepN} \sim 0.1 - 0.2$, H for some ASDEX-U hybrid cases is up to ~ 1.8, and up to ~ 1.6 for H-modes. Such a high H is not found for larger $n_{sepN}$.

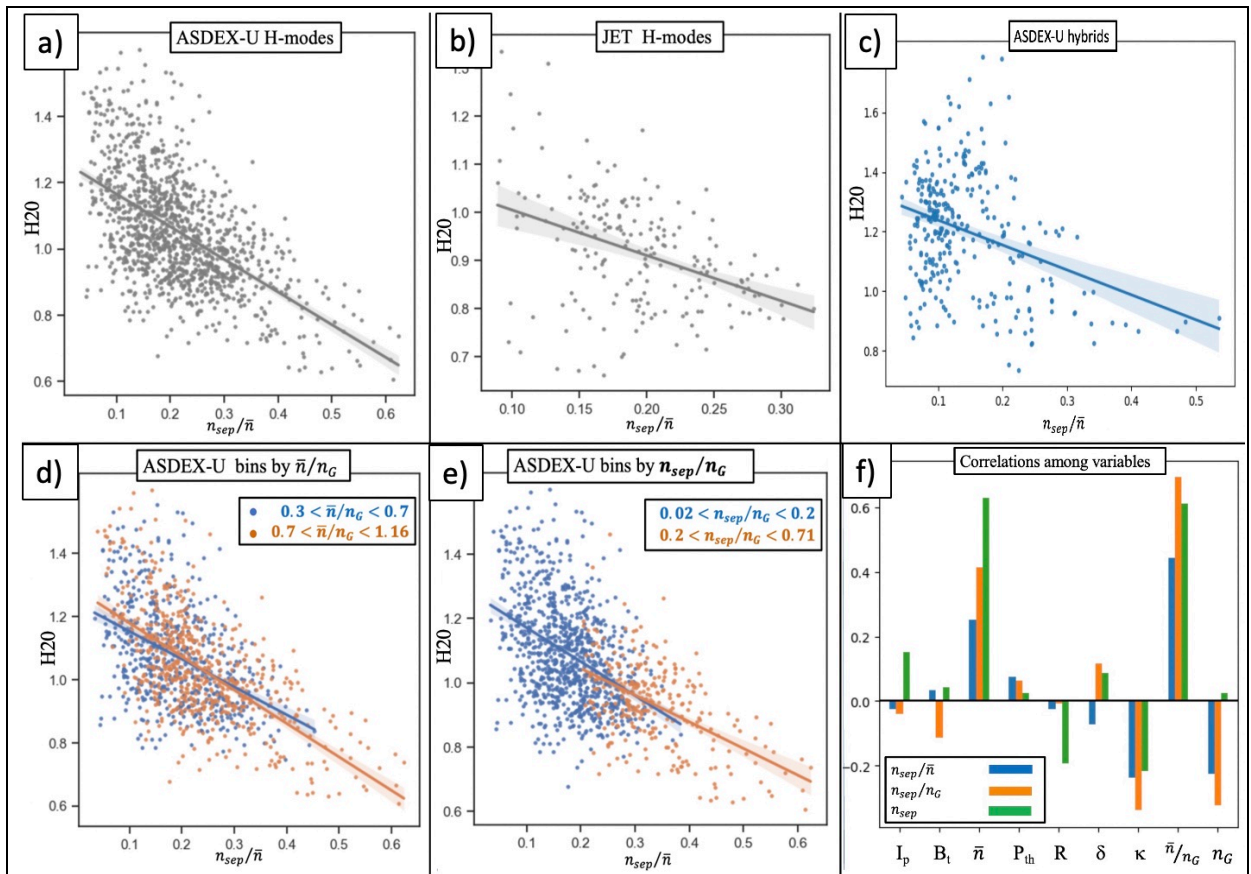

Figure 1 a) Trends of H20 with $n_{sepN}$ for ASDEX-U H-modes. b) H20 vs $n_{sepN}$ for JET-ILW c) H20 vs $n_{sepN}$ for ASDEX-U hybrids. The trends are the same for data binned into two halves by $\bar{n}/n_G$ d) or e) f) $n_{sep}/n_G$. The variable $n_{sepN}$ has low correlations with other independent variables, lower than either $n_{sep}/n_G$ or $n_{sep}$, so it is a new direction in data space.

The trend is not merely a confinement degradation at high values of $n_{sep}$; there is also an equally strong improvement at low values. Trends with $n_{sep}$ have been reported before [1,5,26], but often in terms of $n_{sep}$ normalized by the Greenwald density limit, $n_{sepG} = n_{sep}/n_G$. *The data shows that the effect of $n_{sepN}$ on confinement is not related to proximity to the density limit.* To demonstrate this, we split the data in halves, with a low and a high bin in $\bar{n}/n_G$, and plot the data and regression lines for each. We see in figure 1d that for ASDEX, the lines nearly overlap. If we bin the data into upper and lower values of $n_{sep}/n_G$ the lines are also similar (in fact slightly steeper for lower $n_{sep}/n_G$, figure 1f). Similar results are found for JET-ILW. *The confinement trends, therefore, are not a result of the proximity to $n_G$ of either $\bar{n}$ or $n_{sep}$.*

Hence there is no strong physical reason to normalize $n_{sep}$ by $n_G$. We choose, instead, $n_{sepN} = n_{sep}/\bar{n}$, the normalized variable (available in the database) that is closest to $n_{sep}/n_{ped}$. We remind the reader that $n_{sep}/n_{ped}$ is theoretically expected to affect edge transport barriers, both via gyrokinetic transport (via ETG transport [20,27] or ITG/TEM [25]), and by MHD stability (shifts in the density profile also change $n_{sep}/n_{ped}$).

In order to make sure that $n_{sepN}$ is a good choice for interpreting the data (it is not just corelated to some earlier variable(s) with known trends), we show in figure 1g, the correlations of relevant variables to three density measures containing $n_{sep}$: $n_{sep}$ itself, $n_{sepN}$ and $n_{sepG}$ (for the H-mode data subset that includes $n_{sep}$). The $n_{sepN}$ has lower correlation to most other relevant dependent variables, in particular $\bar{n}$ and $\bar{n}/n_G$. *So $n_{sepN}$ can be considered a new dimension in the data space: trends in it have the least confusion with trends in other standard variables.*

Further statistical tests show that $n_{sepN}$, for this dataset, has slightly greater explanatory power for $\tau_E$ than the usual variables $\bar{n}$ and toroidal field $B_t$, but about 3x less than current $I_p$ and $P_{th}$. In short, it is indeed an important variable, as the visual appearance of figure 1 would indicate. Hence, we should add $n_{sepN}$ to the list of variables characterizing confinement.

Further trends in the data

The trends in $n_{sepN}$ are often stronger at high $\beta_N$, high $\beta_{pol}$ or high Shafranov shift parameter; all these represent total $\beta$ divided by some power of current. High $\beta_{pol}$ operation has been pursued as a route to high confinement on many devices including JT-60-U, DIII-D and EAST [28-32], and it has previously been reported that ASDEX-U and JET show higher H at higher $\beta_{pol}$ and/or $\beta_N$ [33-35]. To describe such trends in this database, we will use a parameter that is closely related to $\beta_{pol}$: the global average of the Shafranov parameter $\alpha$ that is commonly used in local analysis of MHD and gyrokinetics ($\alpha = \epsilon \beta_{pol}$ or equivalently $\alpha = (R/a)\beta\, q_{cyl}^2$, where $q_{cyl} \sim RI_p/a^2 B_t \kappa$ as in [1], and $\beta_{pol} \sim \beta\, \kappa^2/(I_p/a)^2$). For ASDEX-U H-modes, the highest quartile in this parameter shows the strongest trend with $n_{sepN}$ in H-modes, and the lowest quartile shows a weaker trend (see figure 2a). Since there is less data for JET-ILW (figure 2b) and ASDEX-U hybrid modes (figure 2c), we segregate into two bins. The behavior is similar for JET-ILW, while for ASDEX-U hybrids the slopes are similar (within the confidence intervals) but H20 is significantly higher at higher $\alpha$ (or $\beta_N$ or $\beta_{pol}$) together with lower $n_{sepN}$.

We now report a strong dependence of the $n_{sepN}$ trends upon the wall material. The data from ASDEX-U has several different combinations of wall material together with divertor configuration.

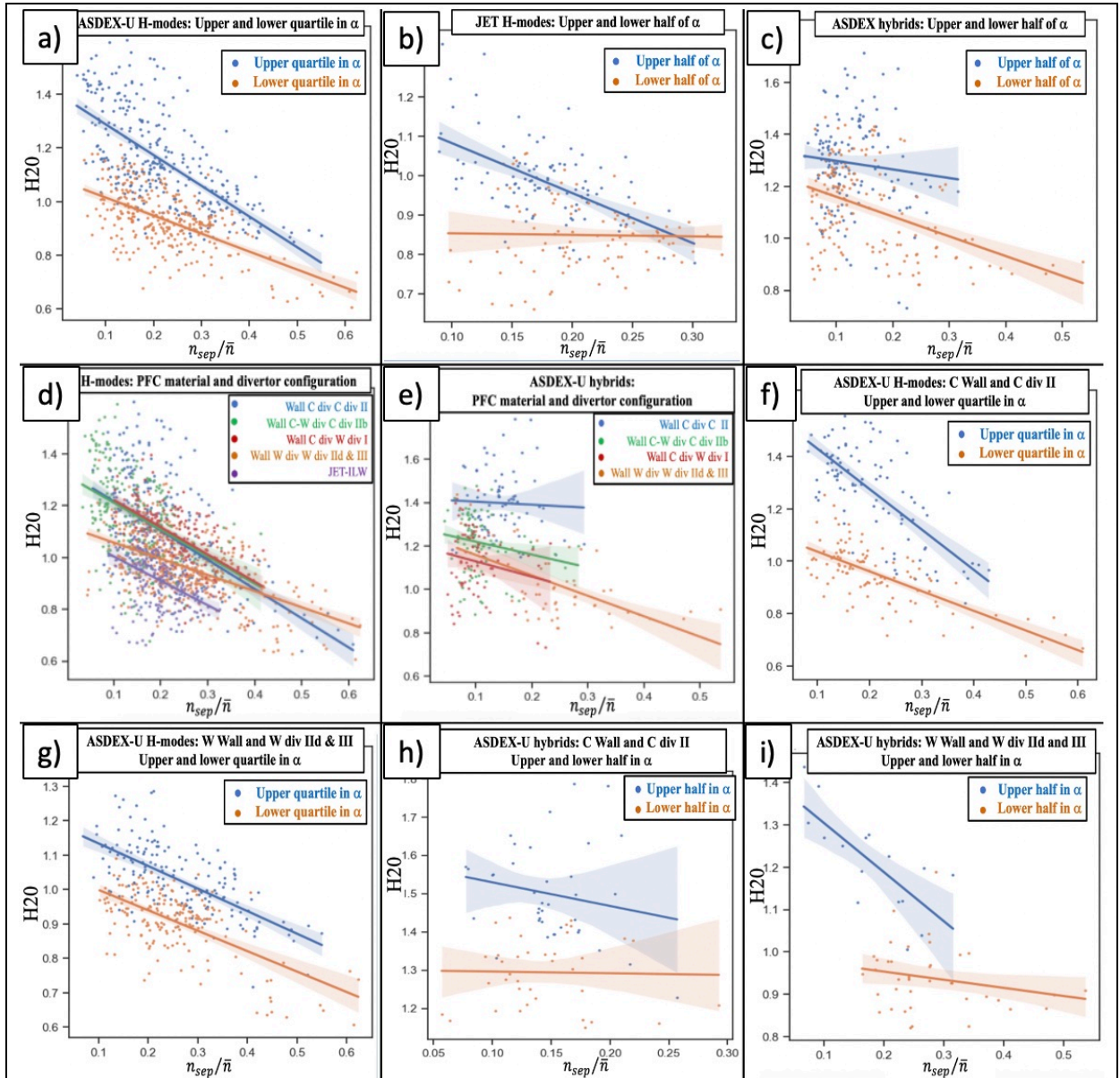

Figure 2 Trends of H with $n_{sepN}$ are sometimes stronger for higher Shafranov shift parameter α (~$\epsilon\beta_{pol}$). a) higher and lower quartiles in α for ASDEX-U H-modes b) higher and lower halves in α for JET-ILW c) higher and lower halves in α for ASDEX-U hybrids. The trends also depend upon wall material and/or divertor configuration. Trends for d) ASDEX-U H-modes e) ASDEX-U hybrids. The wall material also affects whether trends are stronger for higher α : f) ASDEX-U H-modes with C divertor and wall g) ASDEX-U H-modes with W divertor and wall h) ASDEX-U hybrids with C divertor and wall h) ASDEX-U hybrids with W

> divertor and wall. A consistent pattern is: the highest H is always found for the combination of high α together with moderately low $n_{sepN}$.

In figure 2d, we see that for ASDEX-U, the strongest slope in $n_{sepN}$-H plot occurs when the wall and the divertor are both C. The highest H20 is reached for the combination: Carbon (C) divertor material with low $n_{sepN}$ (0.1-0.25). Also, the strength of the trend in $n_{sepN}$ is weakest for all tungsten (W) PFCs. JET-ILW shows a stronger trend, but displaced to lower H20. In Fig 2e we see that for the ASDEX-U hybrids, the highest H (up to ~ 1.8) is found at low $n_{sepN}$ (≲ 0.2) with all C PFCs, and the next highest H are with C divertor and mixed C-W walls. This is similar to the H-modes.

*For every wall and/or divertor type, the highest H always occurs with a conjunction of high α (and thus high $β_{pol}$ or $β_N$) together with fairly low $n_{sepN}$ ($n_{sepN}$ ~0.1 − 0.25). Sometimes there is a different slope of H with $n_{sepN}$ for high α (and $β_N$ or $β_{pol}$), sometimes the slopes are the same. However, the consistent pattern is the one noted at the beginning of this paragraph.*

<u>The nature of the confinement improvement</u>

We now consider the case with the largest confinement improvement: all C PFCs with high α. To reduce changes in other parameters, we window to about the top third of α and also window the data in $I_p$ to allow only about 20% variation ($I_p ∈ (0.79, 1.01) MA$). This also reduces changes in other variables such as $\bar{n}$ and $B_t$, so that variations in H20 likely can be attributed mainly to variations in $n_{sepN}$. We see in figure 3a, an average increase of ~ 40% in H as $n_{sepN}$ goes from ~0.4 to ~0.1. The trend in the confinement time $τ_E = W_{th}/P_{th}$ is shown in figure 3b. To further reduce the effect of residual $I_p$ variations we have normalized by the current, $τ_E/I_p$. This has an enormously stronger increase of ~300%. The magnitude of improvement, apparently due $n_{sepN}$, is rather amazing.

What lies behind this remarkable increase?

Let us plot the thermal stored energy $W_{th}$ normalize by $I_p$ in figure 3c. There is a moderate increase of ~ 30% at low $n_{sepN}$. But the heating power needed to achieve this is reduced by a tremendous amount ~300% (see figure 3d). The ELM type is also indicated in this figure; most are type 1, and a few are type II. Both types arise when the pedestal pressure gradient reaches an ideal MHD instability threshold. Hence, by reducing the $n_{sepN}$, enormously less heating power is necessary to push the pedestal gradients to the MHD limit. *This is prima facie evidence of a reduction of the pedestal transport.* Because of the dependence of the ITPA20-IL scaling law on heating power (~$P_{th}^{-0.64}$) the effect on H is far less than the effect on $τ_E$. Even so, most of the increase in H is due to the reduction in heating power for all C PFCs.

The core of H-modes is stiff [36-43], so we expect the $W_{th}$ to be closely related to the pedestal stored energy and pressure at the top of the pedestal. The thermal $β_{Nth}$ is shown in figure 3e, and it shows a modest increase. The magnitude of change is roughly consistent with calculations in the literature on the effect of shifts of the pedestal density profile that increase the MHD stability threshold [9,12,44]. In particular, there is no indication of a large change in MHD stability which might somehow explain the dramatic reduction in the heating power needed to drive an ELM.

To check that this effect cannot be attributed to changes in other parameters, in figure 3f we plot the trends in the line average density and total magnetic field; they change little. Thus, the predicted L->H threshold power changes little in this $n_{sepN}$ scan ($P_{thresh} \sim \bar{n}^{.72} B_t^{0.8} S^{0.94}$ [45]); the huge variations in power cannot be attributed to expected changes in this threshold. This is qualitatively true of other quantities in the ITPA20-IL scaling law too.

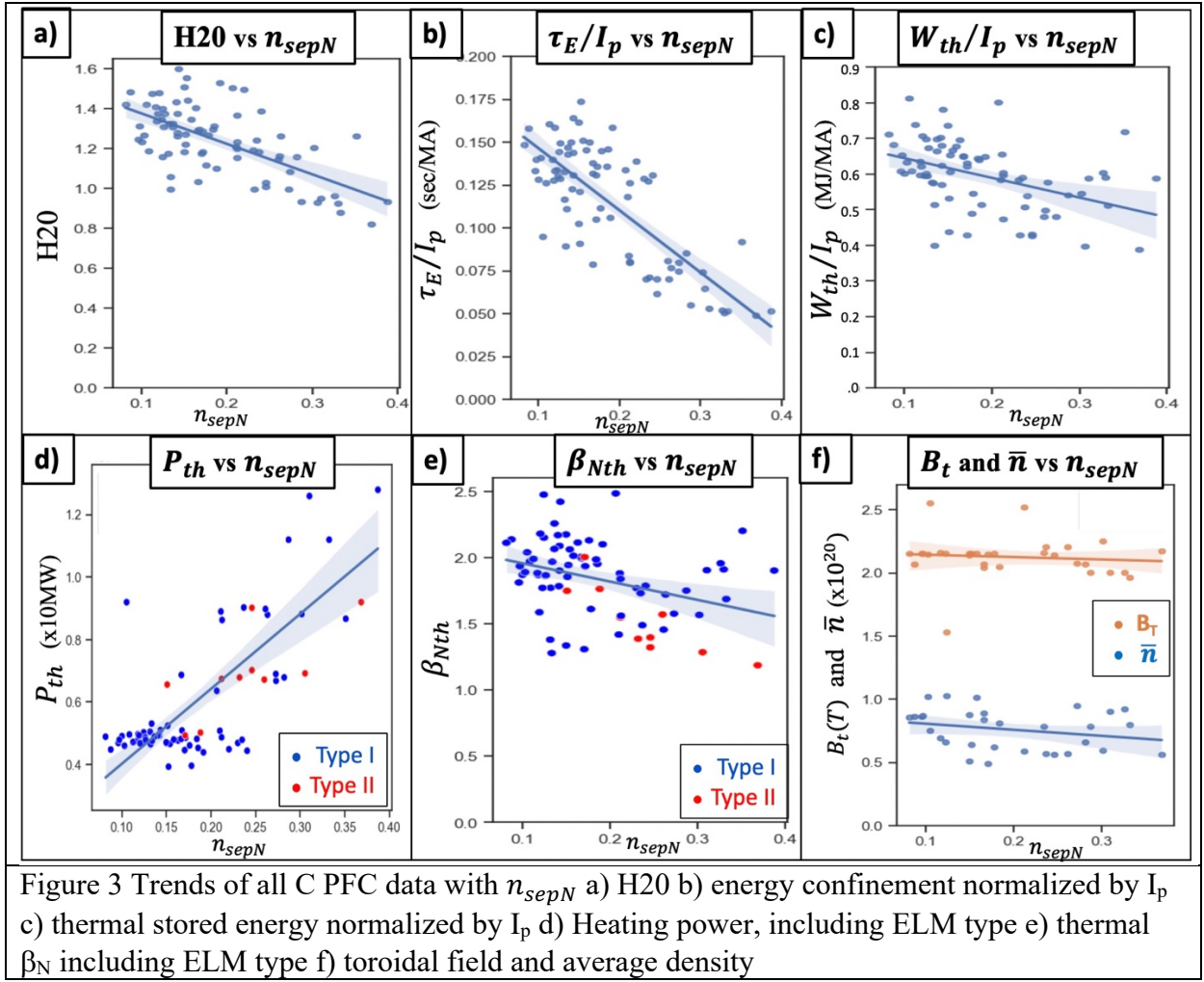

Figure 3 Trends of all C PFC data with $n_{sepN}$ a) H20 b) energy confinement normalized by $I_p$ c) thermal stored energy normalized by $I_p$ d) Heating power, including ELM type e) thermal $\beta_N$ including ELM type f) toroidal field and average density

We now examine whether the core confinement is a substantial contributor to this effect. Fortunately, the ITPA database for all C PFCs has additional data that allows us to check the condition of the core. In Figure 4a, we plot the density peaking $n(0)/\bar{n}$. It does not reach high values, but there is a modest increase ~20% at low $n_{sepN}$. So the several fold decrease in $n_{sepN}$ implies that most of the density profile change occurs in the pedestal, with rather little change in the core. Changes in the pedestal density gradient should not *directly* affect the turbulence determining core profiles, since the latter has a fairly short correlation length. For example, *local* gyrokinetic models show good agreement with H mode profiles. So the fact that the large majority of the change in $n_{sepN}$ reflects a change in the pedestal region reinforces the conclusion that the confinement trends in fig 3 are mainly attributable to that region, which propagates to the core via stiffness.

A significant fraction of the all C PFC data has values for quantities such as $T_i(0)$, $T_e(0)$, and $Z_{eff}$. We have verified that the trends for this subset of the data are exactly the same as in figure 3 above, so it is a representative sample. We examine core quantities such as $T_i(0)/T_e(0)$, the hot particle fraction $W_{fast}/W_{th}$, and the impurity content via $Z_{eff}$ in figure 4b, 4c and 4d, respectively. All these quantities are known to improve ITG/TEM stability and are associated with improved core confinement. *All of these quantities trend oppositely to the direction needed to explain improving confinement as $n_{sepN}$ increases.* This is likely because $P_{th}$ decreases along with $n_{sepN}$, and NBI heating drives $T_i(0)/T_e(0)$ and $W_{fast}/W_{th}$. Reduced heating might also reduce wall sputtering of impurities, decreasing $Z_{eff}$.

Finally, we compute a measure of the pressure peaking which includes the effect of dilution from an assumed carbon impurity $p(0)/<p>$, where $<p> = 2 W_{th}/3\,Vol$ and

$$p(0) = n_e(0)\{T_e(0) + [(7 - Z_{eff})/6]T_i(0)\} \qquad \text{eq(1)}$$

For this approximate measure, we must assume that $Z_{eff}$ is constant since the 0D database has no information regarding its profile. As we see in figure 4e the pressure peaking is essentially statistically constant, and shows no strong trend as $n_{sepN}$ decreases.

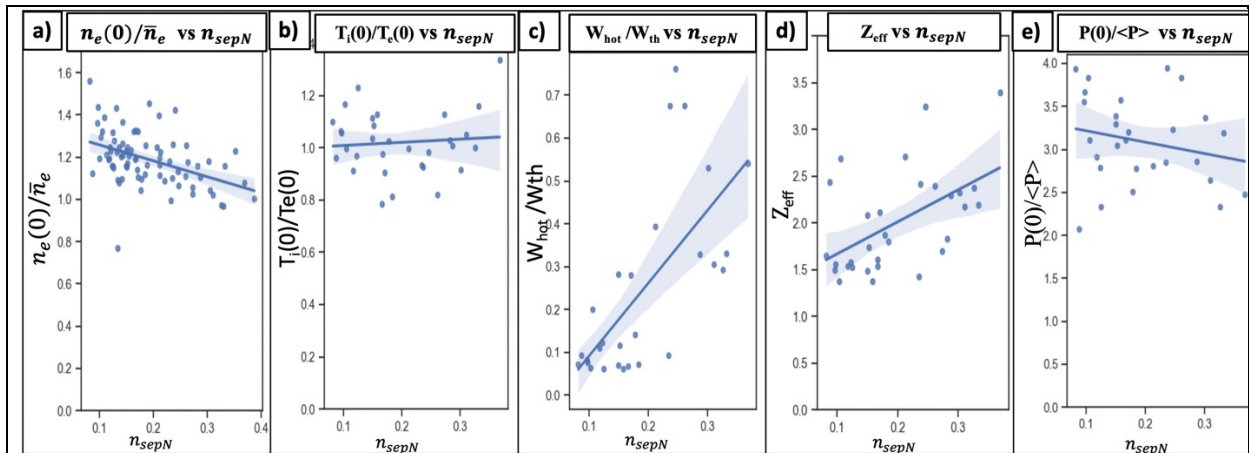

Figure 4 Trends of core related quantities with $n_{sepN}$ for all C PFCs a) density peaking b) The ratio of ion to electron central temperature c) the ratio of hot stored energy to thermal stored energy d) the effective charge e) a measure of pressure peaking eq(1). These all indicate the the confinement trends in figure 3 cannot be explained by core physics, but must result fro pedestal physics.

Core confinement seems not to be responsible for the trends in figure 3. This reinforces the conclusion that the trends arise from the pedestal.

To confirm this, multiple analysis of H-mode discharges find that the core stored energy is closely related to the pedestal stored on ASDEX-U and JET-ILW [36-40]. Data for ASDEX-U and JET show that the stored energy or $\beta_{pol}$ are roughly a constant times the corresponding pedestal quantity. In addition, the temperature profiles are stiff in the T scale length. With weak trends in density peaking, this stiffness reinforces the conclusion that $W_{th}/W_{ped}$ has little variation. This qualitative behavior is also a consistent feature of gyrokinetic transport models based upon ab initio simulations that show good agreement to experiments [41-43].

All lines of evidence point toward the same conclusion: the overall boost in confinement (Figure 3) stems from an improved pedestal. A strong reduction in pedestal transport as $n_{sepN}$ decreases implies that less heating power is needed to reach about the same pedestal pressure (which is MHD limited). The core pressure reflects the pedestal pressure through stiffness.

For other PFC combinations, a carbon divertor with mixed C-W walls has behavior similar to all carbon, but the trends are slightly weaker. The trends for all W PFCs are, however, considerably weaker than for all C, including for this windowed dataset. We compare the all C and all W wall types for $\tau_E$, $W_{th}/I_p$ and $P_{th}$ in figure 5a, 5b and 5c, respectively. The stored energy is slightly lower for W, but the heating power is higher and does not decrease nearly as much at low $n_{sepN}$. The trends in the stored energy are consistent with an improvement in pedestal MHD stability with a shift in density profile, as found previously in this range of $n_{sepN}$ [9, 12, 44].

What is the reason for this weaker trend with W PFCs? A widely discussed aspect of W PFCs is the need for higher fueling to avoid unacceptable W radiation in the core, and the deleterious impact of this upon confinement. This database has evidence of this as well, which bears on the trends in $n_{sepN}$. The largest differences between W and C PFCs are at small $n_{sepN} \lesssim 0.2$; the power reduction with low $n_{sepN}$ for W is much less than C (See figure 5a-c). To see if W and C PFCs are similar with similar fuel rates, we window it to values corresponding to roughly in the midrange for W PFCs, and where there are some values in the data for C PFCs (fueling ∈ (1.2,2.4) $10^{21}\ sec^{-1}$). Then the trends are much closer as seen in figure 5d. This is primarily because the low $n_{sepN}$ region of the data is mostly absent: fueling tends to preferentially raise $n_{sep}$ relative to $\bar{n}$ so there are no cases with low $n_{sepN}$. To examine evidence in the database regarding the importance of higher radiation from W at low $n_{sep}$, we subtract radiation from the heating power, defining $\tau_{E\ rad} = W_{th}/(P_{th} - P_{rad})$. (This quantity can only be considered for qualitative purposes, since a significant portion of $P_{rad}$ often arises outside of the separatrix. However, at lower $n_{sep}$, this fraction is likely considerably smaller.) The trends in $\tau_{E\ rad}$ are, then, significantly closer for these two choices of PFCs (compare figure 5e to 5a). After windowing in fueling as before, the trends overlap (see Figure 5f). So in this data, fueling and core radiation are major factors in the difference between C and W PFCs. This is not a full explanation of all the data, however. If we window to low fueling ( $< 0.25 \times 10^{21}\ sec^{-1}$, see figure 5g), there are large differences between all C and all W PFCs. These are considerably reduced when accounting for radiation using $\tau_{E\ rad}$, but differences remain (see figure 5h). This might be related to differences in the pedestal density profile that are not described by $n_{sepN}$, that might arise from the different neutral reflection of W PFCs compared to C PFCs. [46]. Another factor might be that core heating often includes additional electron heating for such cases [47]. We further discuss the PFC differences in the interpretation section.

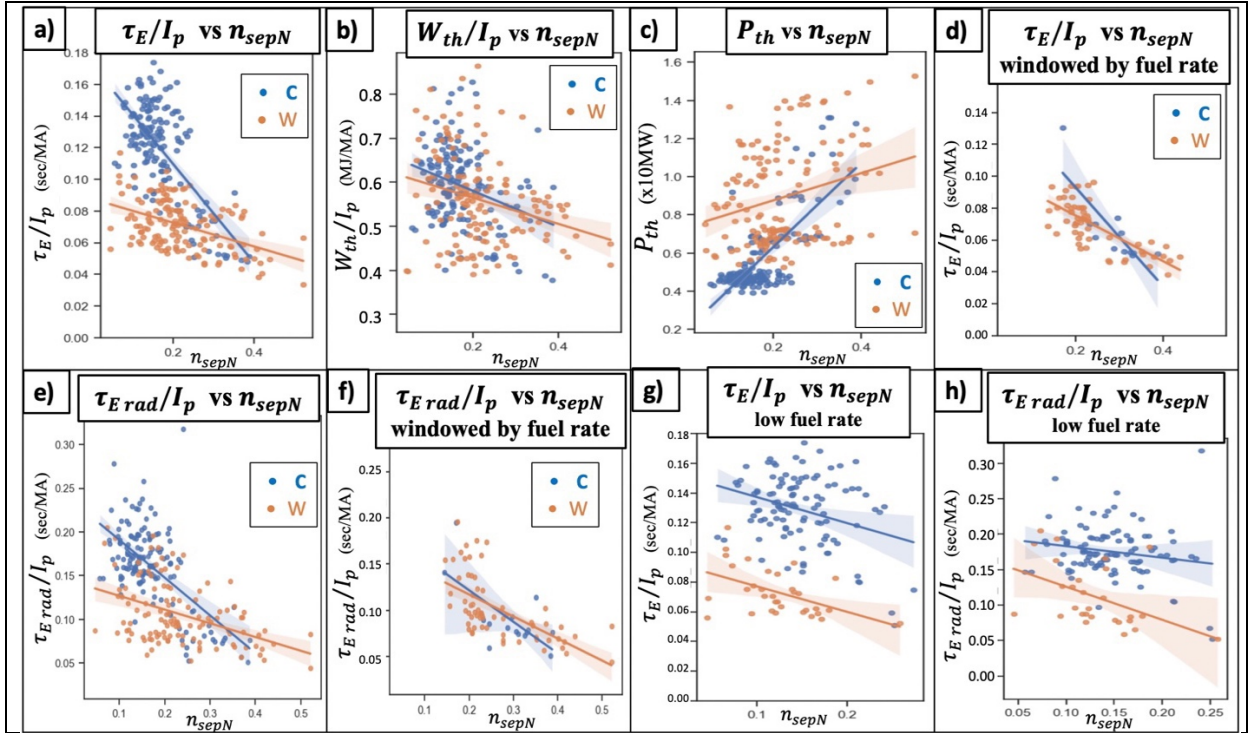

Figure 5 Comparison of trends versus $n_{sepN}$ for all C PFCs with all W PFCs a) confinement time normalized by $I_p$ b) thermal stored b) stored thermal energy normalized by $I_p$ c) heating power. The heating power reduction at low $n_{sepN}$ is much weaker for all W PFCs d) normalized energy confinement, for data windowed by fueling rate typical of W, brings the trends closer e) subtracting the radiation power from heating power also brings the trends closer f) doing this plus windowing in fuel rate, the W and C PFC cases are statistically the same g) differences for low fueling rate are quite large h) subtracting the radiation power from heating power brings the trends closer, but significant differences remain.

Hybrids (Improved H-modes) and unwindowed data

Here we include datasets of the hybrid and standard H-modes together. Up to now we have considered the variable $n_{sepN} = n_{sep}/\bar{n}$, which can be said to describe density peaking mainly coming from the pedestal region. But hybrids have significant peaking in the core (see fig 6a) as well. Previous analysis have found that such core peaking is significant for higher confinement in ASDEX-U [48, 49]. We find that $n_{sepN0} = n_{sep}/n(0)$, the inverse of the *total density peaking from the axis to the separatrix, then, becomes the variable that best explains the data.*

To isolate the effects of density peaking (measured by $n_{sepN0}$), we go through the following steps. We, first, consider all C PFCs, which have the strongest variation in H20 (especially when hybrids are included). To minimize the interference from the strong effects of α variations, we bin to even higher α (α > 0.47). We also window in current as before. This high α bin includes the highest confinement cases.

There is a clear trend of H20 with $n_{sepN0}$ (see Figure 6b). Now consider linear regressions to see if this variable best explains the H20 variation. As seen in table 1a, it has the greatest explanatory power (as indicated by the coefficient of determination $R^2$, which can be interpreted as the fraction of the variation in the data can be explained by the variable). *In fact, the single variable $n_{sepN0}$ performs about as well as a bivariate regression in both the pedestal variable $n_{sepN}$ and the core density peaking $\bar{n}/n(0)$.* In other words, the single variable $n_{sepN0}$ describes the H20 variation as well as both the two variables for pedestal and core density peaking. The hybrids have the highest confinement, and they also cluster at the highest total density peaking (since they combine core density peaking with pedestal density peaking). This variable has better explanatory power than other variables that have sometimes been suggested as responsible for high confinement: low $v^*_{avg}$, high $T_i(0)/T_e(0)$, low $\bar{n}$, whether the shot was a "hybrid", and δ (triangularity). (Some details: the regressions are linear rather than log-linear since this usually led to the highest $R^2$. The average collisionality variable $v^*_{avg}$ is defined in (eq(1c) of reference [1]. We define the "hybrid" variable to be 0 for standard H-modes and 1 for hybrids.)

Now we return to the weaker windowing in α (α > 0.41, as used in the previous section) but include hybrids in the data set. We consider several regressions including the core density peaking in Table 1b. We see that both $n_{sepN}$ and $\bar{n}/n(0)$ can, individually, explain some of the variation in the data, and bivariate regression, improves $R^2$. *But again, the single variable $n_{sepN0}$ explains the variation virtually as well as these two variables.*

To see which *pair* of variables best explains the variation in H20, consider bivariate regressions with $n_{sepN0}$ with an additional variable (See Table 1b). The pair $n_{sepN0}$ and α is the best. Even though the windowing reduces the range in α, it still emerges as having the highest explanatory power.

All the regressions give improvement in H with lower $n_{sepN0}$- they have to, in view of the ubiquitous nature of this in the data. The following figures are a visual representation of the quality of the fit: how well these two variables explain most of the variation of the data. This visual presentation is perhaps more familiar than the metric R2. Compared to our single variable $n_{sepN}$ (invoked in the previous section (fig 6c)), the scatter is very considerably reduced (see figure 6d) by the pair $n_{sepN0}$ and α.

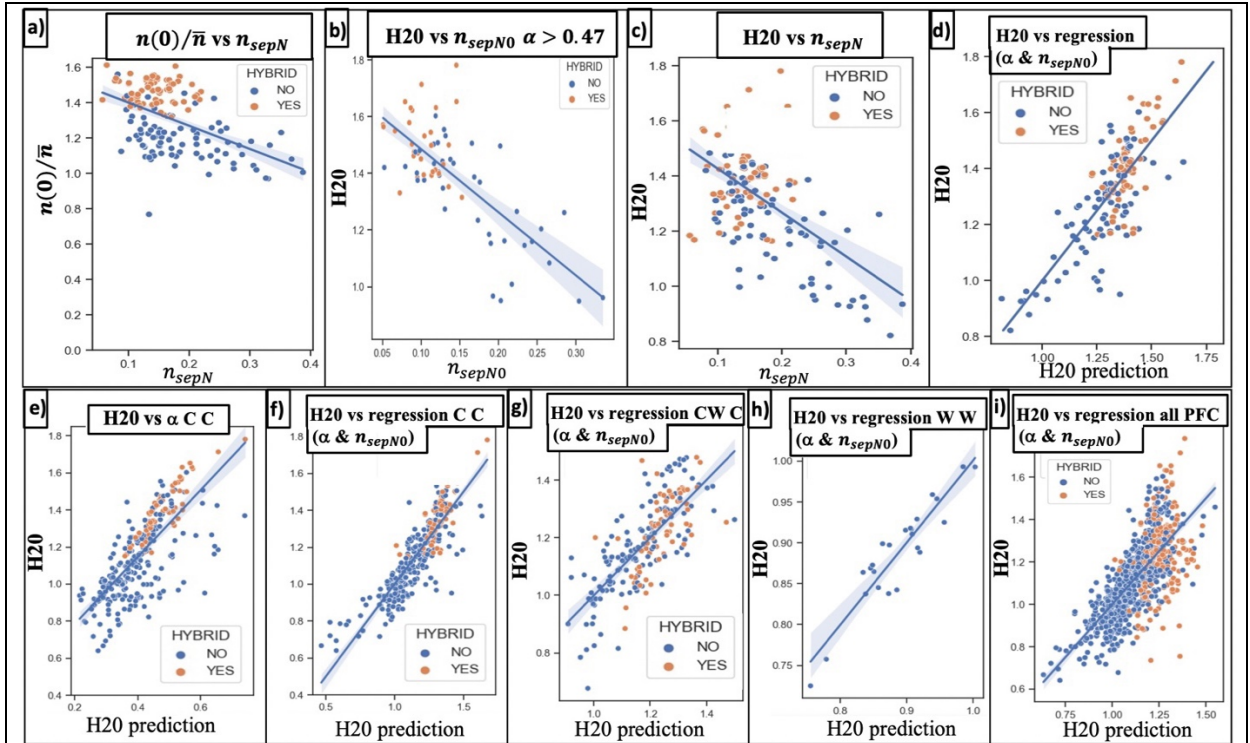

Figure 6 a) density peaking is higher for all C hybrids b) windowed for high α, there is a strong trend of H20 with $n_{sepN0}$ c) for less windowing in α (α > 0.41), the trend in $n_{sepN}$ is clear but has considerably more scatter d) A bivariate regression in $n_{sepN0}$ and α explains the data best, and has considerably reduced scatter e) for data of all C without any windowing, α explains the data best, but there is considerable scatter f) a bivariate regression in α and $n_{sepN0}$ explains the data best and has remarkably little scatter g) the same two variables explain mostg of the variation for C divertor PFCs but mixed W and C main chamber walls h) these two variables explain the limited all W dataset that includes n(0) quite well i) they also explain about half the variation in the full dataset with all wall types; scatter is increased because PFC type is a significant variable.

Next we consider the dataset of all C PFCs with no windowing in α or other variables. See table 1c. The single variable with highest $R^2$ is α, but there is nonetheless considerable scatter (figure 6e). A trend of increasing H factor with $\beta_N$ has been noted before for ASDEX-U (34), and improved confinement at high $\beta_{pol}$ has been found on many devices. We now combine α with one additional variable (bivariate regressions). Once again, the best pair is α and $n_{sepN0}$. The coefficient of determination $R^2$ is quite high, 0.74, and there is much less scatter (see figure 6e) than for α alone. The trends in H20 are rather well explained by just these two variables. (Note that the next best pair is α and $n_{sepN}$, the pedestal density peaking.) We have considered tri-variate regressions with these two variables and one of the others. The best additional variable is "hybrid", closely followed by $\nu^*_{avg}$, but adding these variables gives only a marginal improvement (increase of $R^2$ by ~ 0.03) which is hardly visible in graphs like these. (The relatively low importance of being a hybrid agrees with other analysis in the literature that suggest that they are similar to other H-modes [37].)

In conclusion, the variables α and $n_{sepN0}$ explain most of the variation of H20 for all C PFCs. The very high H20 factors of hybrid cases (up to ~ 1.8) can almost entirely be explained by the combination of high α and low $n_{sepN0}$, i.e., a high total density peaking from separatrix to axis.

Table 1: regression results to identify the variables with the highest explanatory power

| a) α > 0.47 $I_p \in (0.79, 1.01)$ MA | | b) α > 0.41 $I_p \in (0.79, 1.01)$ MA | | c) All C PFC data | |
|---|---|---|---|---|---|
| Regression variable(s) | $R^2$ | Regression variable(s) | $R^2$ | Regression variable(s) | $R^2$ |
| $n_{sepN0}$ | 0.547 | $n_{sepN}$ | 0.308 | α | 0.442 |
| $n_{sepN}$, $\bar{n}/n(0)$ | 0.54 | $\bar{n}/n(0)$ | 0.305 | α, $n_{sepN0}$ | 0.736 |
| $n_{sepN}$ | 0.467 | $n_{sepN}$, $\bar{n}/n(0)$ | 0.420 | α, $n_{sepN}$ | 0.704 |
| $v^*_{avg}$ | 0.437 | $n_{sepN0}$ | 0.409 | α, $\bar{n}/n(0)$ | 0.654 |
| $\bar{n}/n(0)$ | 0.344 | | | α, $v^*_{avg}$ | 0.652 |
| δ (triangularity) | 0.281 | $n_{sepN0}$, α | 0.624 | α, "Hybrid" | 0.593 |
| "Hybrid" | 0.274 | $n_{sepN0}$, $v^*_{avg}$ | 0.503 | | |
| $T_i(0)/T_e(0)$ | 0.195 | $n_{sepN0}$, "Hybrid" | 0.438 | | |
| $\bar{n}$ | 0.068 | $n_{sepN0}$, δ | 0.433 | | |

We now consider these same two variables for other PFC types on ASDEX-U. (Note there is no data for n(0) for JET-ILW). The fit is also reasonably good for C divertors and mixed C and W walls (figure 6f). Results for these and other wall types are given in Table 4. For all W walls, only a very limited number of shots in the database have n(0), but the bivariate regression fits the data very well (figure 6g); however the dependence is almost entirely in α. Only in the case of C walls and W divertor does bivariate regression in these variables explain significantly less than half the variation. Note that this older divertor geometry is very different from all other cases, and is very open. Substituting $v^*_{avg}$ in for $n_{sepN0}$ markedly improves $R^2$, but still only to 0.464. Including $v^*_{avg}$ in a trivariate regression for all the other PFC types in table 3 gives only rather small improvement in $R^2$ (≲ 0.05, and hardly visible).

Note that the difference in many regression coefficients for various wall types are larger than the error bars (See table 2). *Hence, the PFC type (or divertor geometry) is again a significant variable, as found in the preceding sections.* Nonetheless the regression coefficients have roughly similar magnitude (except for all W PFCs).

If we consider a bivariate regression of α and $n_{sepN0}$, for all PFC types together (hence ignoring PFC type), the $R^2$ is significantly less than for many wall types individually (scatter is greater). This is not surprising since the wall type is a significant variable which is best not ignored. See figure 6h and Table 2. A tri-variate regression including $v^*_{avg}$ only modestly improves $R^2$ (from 0.500 to 0.539).

Table 2: coefficients of the regression $H20=C_0+C_1\alpha+C_2 n_{sepN0}+C_3 v^*_{avg}$

| PFC type | $R^2$ | $C_0$ | $C_1$ | $C_2$ | $C_3$ |
|---|---|---|---|---|---|
| Wall C div C | 0.736 | $0.901 \pm 0.03$ | $1.154 \pm 0.05$ | $-1.344 \pm 0.07$ | 0 |
| Wall CW div C | 0.545 | $0.897 \pm 0.03$ | $0.972 \pm 0.07$ | $-1.275 \pm 0.11$ | 0 |
| Wall W div W | 0.865 | $0.168 \pm 0.06$ | $2.432 \pm 0.2$ | $-0.081 \pm 0.12$ | 0 |
| Wall W div CW | 0.319 | $1.0313 \pm 0.05$ | $0.644 \pm 0.1$ | $-0.876 \pm 0.1$ | 0 |
| Wall C div W | 0.464 | $0.917 \pm 0.02$ | $0.544 \pm 0.09$ | 0 | $-0.556 \pm 0.08$ |
| All types | 0.500 | $0.957 \pm 0.02$ | $0.760 \pm 0.04$ | $-0.926 \pm 0.05$ | 0 |
| All types | 0.539 | $1.001 \pm 0.02$ | $0.670 \pm 0.04$ | $-0.646 \pm 0.06$ | $-0.142 \pm 0.03$ |

Interpretation of the results

The variable pair $n_{sepN0} = n_{sep}/n(0)$ and $\alpha$ exhibit the most explanatory power to order and understand the experimental trends in H20 variation. For all C PFCs, the combination explains 74% of the variation of H20 in the data (figure 6f). This qualitatively agrees with many previous results in the literature, on many devices. In this section we examine

1) the strong connection of this finding to experimental results on other devices, and to the importance of PFC material upon confinement

2) the strong connection of the trends found in the dataset to the general predictions of gyrokinetic simulations. Detailed recent analysis/simulations show that, because of a dynamical constraint in the gyrokinetic system, instability access to free energy from temperature gradients can be prevented by a under conditions of *strong density gradients*, for magnetic geometry which in tokamaks occurs at high $\alpha$ or high $\beta_{pol}$.

3) The extension of the methodology - using the same pair of variables- to explain discharges with internal transport barriers -ITBs.

We begin with establishing experimental connections.

Improved confinement with higher $\beta_{pol}$ (or $\beta_N$) has been found on JT-60U, ASDEX-U, DIII-D, EAST and JET-ILW [28-35]. This is sometimes associated with an ITB, and sometimes not. In the later part of this section, we discuss cases with ITBs on devices other than ASDEX-U, and find that the two variables $n_{sepN0}$ and $\alpha$ seem quite relevant.

Many devices have also reported improved confinement with lower recycling [4-8, 13-16]. Recycling should be highly correlated to $n_{sep}$. It is very plausible that the separatrix density is the route by which recycling effects the core transport- namely, by affecting this boundary condition of the core.

Many devices also report improved confinement with *core* density peaking [31,48-51]. To quote a 2018 review of tokamaks and stellarators by F. Wagner, discoverer of the H-mode, "Some of the essential preconditions to improved confinement regimes are surprisingly identical for tokamaks and stellarators: recycling control and avoiding gas puffing, sufficient heating power to overcome thresholds and peaking of the density profile" [51]. The total density peaking $n_{sepN0}^{-1} = n(0)/n_{sep}$ is a single parameter that includes both the core $n(0)/n_{ped}$ and the pedestal peaking $n_{ped}/n_{sep}$. The correlations found here indicate that the quote above might be extended to this total density peaking. Core density peaking can occur together with ITBs, and can be beneficial or essential for them [25,31,52,53]. We will see that $n_{sepN0}$ appears correlated with H20 for a few ITB cases examined below.

The PFC material is also found to be a significant variable for ASDEX-U. The importance of $n_{sepN0}$ for all C PFCs implies an explanation for the lower confinement of high Z PFCs in the database. It is well known that high density peaking leads to high Z accumulation due to neoclassical transport. In such a case, high transport from a source other than neoclassical is needed to flush the W out (e.g. instabilities). *The results above find that the regimes of highly elevated H20 are engendered by precisely the conditions that lead to unacceptable W accumulation- density peaking (in the pedestal and/or the core) giving strong inward W flux, together with low instability transport*. Given this contradiction, operational measures to attempt to avoid high Z accumulation will either 1) lead to additional transport, of one kind or another, that flushes W out but also requires higher heating power and thus lowers confinement, or 2) suffers from highly reduced transport power due to high core radiation, or 3) limits the density peaking itself, and hence the confinement.

Furthermore, low $n_{sep}$ tends to create SOL conditions with high W sputtering and hence a high W source to the plasma. Operation to limit this W source (e.g. gas puffing) tends to increase $n_{sep}$ and hence increase $n_{sepN0}$, and thus, reduces H20.

Qualitatively then, even if low $n_{sepN0}$ is attained with W PFCs, it must go along with other operational measures that increase transport (e.g., gas puffing, central electron heating, etc.), or, suffer strong W radiation that reduces the effective heating power (hence confinement). *Thus, the H20 trends at lower $n_{sepN0}$ should be expected to be less favorable for high Z PFCs like W than for low Z PFCs like C.*

The features, delineated above, are qualitatively consistent with oft mentioned characteristics of operation of ASDEX-U and JET after adopting W PFCs, especially in the divertor. What is new is the particularly strong link, found in the data for low Z PFCs, between the total density peaking $n_{sepN0}$, high H20 and low transport, which provides a concise explanation for the observations.

We now examine an underlying theoretical basis to understand the trends in confinement in terms of the pair of variables: $n_{sepN0}$ and $\alpha$. It is interesting that a very recent paper (see ref [25]) demonstrates, via basic theoretical analysis aided by extensive gyrokinetic simulations, the fundamental and oversize role that (logarithmic) density gradients ($n_{sepN0}$ is a measure) can play in boosting confinement. The active mechanism responsible for this phenomenon is the existence of a basic constraint in the gyrokinetic system (the charge weighted turbulent particle flux must vanish) that can prevent turbulence from accessing free energy in equilibrium gradients- suppressed turbulent transport. Improved confinement is the consequence.

The constraint is important in circumstances where there are significant density gradients together with magnetic geometries where trapped electrons tend to decouple from the fluctuations. In tokamaks, such geometries arise at high local $\alpha$ or, when magnetic shear $\hat{s}$ is small or negative. Higher $\beta_{pol}$ increases $\alpha$, and also gives stronger bootstrap currents that reduce $\hat{s}$. So basic characteristics of the gyrokinetic system imply that turbulent transport will be reduced in circumstances with, simultaneously, high density gradient and high $\beta_{pol}$ (or high $\alpha$). The relevant density gradient is a logarithmic one $\sim (1/n)\,dn/dr$, and the profile averaged value of this in the closed field line region is determined precisely by the quantity $n_{sepN0} = n_{sep}/n(0)$. (Note that reversed shear obtained transiently, without high $\beta_{pol}$, is similarly efficacious to $\alpha$.)

*In terms of the 0D variables available in this database, this theoretically predicted turbulence reduction would be expected to depend upon $n_{sepN0}$ and $\alpha$ -precisely the variables found to be most important in the ASDEX-U data.* It is remarkable that the basic theoretical/computational physics and the experiment are converging to isolate similar variables, that best determine plasma confinement.

This constraint-based theory applies to both pedestals and ITBs (and core profiles without ITBs as well, if the local $\alpha$ is large enough or $\hat{s}$ small enough). Confinement could be improved by a combination of reducing pedestal and/or core transport. Hence, the importance of low $n_{sepN0}$. As a very preliminary examination of the applicability of these ideas to global confinement with ITBs, we have found several published cases for high $\beta_{pol}$ discharges on DIII-D [54] [31] published data to allow approximate evaluation of $\alpha$, $n_{sepN0}$ and an H factor. See Appendix 1. *The cases with H ~ 1.4-1.5 all have high $\alpha$ and low $n_{sepn0}$ in the same range where ASDEX-U finds comparably elevated H20 values.* In fact, using the regression for all C PFCs from Table 2, the predicted H factors for these DIII-D discharges with ITBs are remarkably close to the experimental values for these ITB discharges, *despite the fact that the regression is solely from ASDEX-U data.* Predictions are roughly within ~ 10% -comparable to the accuracy for the ASDEX data from which the fit came. Gyrokinetic analysis of some of ITB discharges has highlighted the importance of density peaking in the core [31]. This preliminary investigation motivates a more extensive analysis of 0D experimental ITB data with larger sample sizes. This, of course, would optimally be done in conjunction with gyrokinetic profile analysis.

Finally, a very different but equally significant direction of investigation is also motivated by these results. To extrapolate the highest H factors to burning plasma regimes, it would be very beneficial to have a PFC material that is more akin to C than to W. But it is well known that for conventional solid PFCs, only high Z materials have the requisite lifetime to be practical. However Liquid Metal (LM) PFCs that can be replenished are a potential option. High Z LMs like tin have been considered for use as PFCs, because they are capable of high temperature operation like W (surface temperature $\gtrsim 1000°C$). However, these would have similar core impurity issues to W. Lithium is one possibility being considered in the community, but its high vapor pressure presents some challenges. Recent advances [55] indicate that there are suitable LM alloys with the high operating temperature like W, and low Z sputtering akin to C. Such materials could be very beneficial for extending the elevated H20 results in todays' devices with C PFCs to burning plasma operation.

Conclusions

We summarize our findings of a somewhat extensive analysis of the ITPA database carried out in the backdrop of a theoretical framework for confinement associated with transport barriers.

1) Two fundamental determinants of the confinement enhancement H20 turn out to be, the ratio $n_{sepN} = n_{sep}/n_{ped}$ and the PFC materials. With stiff core transport (where core profiles are strongly dependent upon the pedestal top) it follows that the confinement time $\tau_{th} = W_{th}/P_{th}$ is just as sensitive to the heating power needed to reach the pedestal MHD stability limit as it is to the value of that pedestal pressure limit [24].

2) The data indicates that the required heating power, strongly dependent upon the separatrix and SOL conditions, can vary by up to a factor of ~ three. Lower heating power for the same pedestal pressure is due to reduction in the pedestal transport. In fact, recent extensive gyrokinetic simulations show precisely such a dependence of pedestal transport on $n_{sep}/n_{ped}$ ; it is lowered as this ratio goes down. Thus, lower $n_{sep}/n_{ped}$ boosts confinement, most likely, via suppressing pedestal (or ITB) transport

3) For C PFCs in the divertor, though most of the improvement in confinement time is due to reduced transport in the pedestal, improvements in MHD stability may also play a role; the latter may be the dominant effect for W PFCs in the divertor.

4) For cases with significant core density peaking (e.g. improved H-modes), the variable $n_{sepN0} = n_{sep}/n(0)$, the total density peaking across the closed field line region, is found to be more suitable for data interpretation. The density ratio, and the global average of the Shafranov parameter α used in local gyrokinetic and MHD calculations, constitute the most important variables for explaining trends in H20. Since $\alpha = \epsilon\beta_{pol}$, this agrees with many experimental findings that high $\beta_{pol}$ improves confinement. It is also consistent with many experimental findings that low recycling improves confinement, since low recycling should give low $n_{sepN}$ or $n_{sepN0}$. Using regressions with both of these variables, most of the variation in H20 can be explained (up to ~ 74%) for all C PFCs.

5) The differences between C PFCs and W PFCs likely stems from the combination of 1) the strong importance of total density peaking to reaching high H20 where transport is low, and 2) the known tendency of W to accumulate unacceptably under conditions of strong density peaking and low instability driven transport.

6) As emphasized earlier, the fact that both recent advanced theoretical/computational studies and extensive data analysis lead to valorizing the two variables: α and $n_{sepN0}$ as strong determinants of confinement (quality of the transport barriers, pedestals or ITBs), is extremely promising news for understanding the physics of confinement. This could be useful to aid the design of future machines and may point the direction to new ways to optimize confinement. Preliminary investigation of ITB cases on DIII-D finds that the trends in H20 are also consistent with the importance of these variables.
7) These results motivation the development of PFC materials with low Z sputtering that are optimally compatible with operation in burning plasmas. These would need to be replenished (due to their high erosion), so liquid metals are one logical choice. Recent advances indicate that there may be multiple novel alloy possibilities for this [55].

Acknowledgements

This work is supported by ExoFusion and by U.S. DOE under Grant Nos. DE-FG02-97ER54415 and DE-SC0022154. We gratefully acknowledge F. Ryter and G. Verdoolaege for assistance with database related matters, and S. Ding for the DIII-D parameters.

Appendix 1

Table A1: high $\beta_{pol}$ DIII-D discharges from the literature with both ITBs and ETBs

| Shot# and (time) | α | $n_{sepn0}$ | $H_{pred}$ | $H_{98}$ |
|---|---|---|---|---|
| 180257 (3700-3800ms) | 0.73 | 0.25 | 1.41 | 1.56 |
| 180257 (4720-4820ms) | 0.72 | 0.20 | 1.46 | 1.51 |
| 190904 (2800) | 0.74 | 0.32 | 1.32 | 1.19 |
| 190904 (3000) | 0.74 | 0.29 | 1.37 | 1.20 |
| 190904 (3800) | 0.81 | 0.33 | 1.39 | 1.38 |
| 190904 (4800) | 0.93 | 0.23 | 1.67 | 1.52 |

Data taken from [31,54]. Note that the errors in the prediction are ~ 10% or less, despite the regression coming exclusively from ASDEX-U data.